\documentclass[a4paper,english,aps,prl,twocolumn,floatfix,showpacs,amsfonts,amssymb,superscriptaddress]{revtex4} 
\usepackage[T1]{fontenc}
\usepackage[latin1]{inputenc}
\usepackage{graphics}
\usepackage{amssymb}
\usepackage{amsmath}
\usepackage{overpic}

\newcommand{\be}{\begin{equation}}
\newcommand{\ee}{\end{equation}}
\newcommand{\bea}{\begin{eqnarray}}
\newcommand{\eea}{\end{eqnarray}}

\newcommand{\lasco}{La$_{2-x}$Sr$_{x}$CuO$_{4}$}
\newcommand{\ybco}{YBa$_{2}$Cu$_{3}$O$_{6+x}$}
\newcommand{\bscco}{Bi$_{2}$Sr$_{2}$CaCu$_{2}$O$_{8}$}

\def\a{\alpha}
\def\b{\beta}
\def\e{\varepsilon}
\def\d{\delta}
\def\g{\gamma}

\def\l{\lambda}

\def\n{\nu}

\def\s{\sigma}

\def\D{\Delta}

\def\ra{\rightarrow}
\def\up{\uparrow}

\def\down{\downarrow}

\def\bk{{\bf k}}

\def\bq{{\bf q}}

\def\DD{{\cal{D}}}

\def\nn{\nonumber}
\def\lb{\label}
\def\pref#1{(\ref{#1})}

\newcount\bozza \bozza=0
\ifnum\bozza=1
\newdimen\shift \shift=-2truecm
\def\lb#1{%
{\label{#1}\rlap{\kern\shift{$\scriptstyle#1$}}}}
\else\def\lb#1{\label{#1}} \fi

\begin{document} 
\title{Fluctuation conductivity from interband pairing in pnictides}
\author{L. Fanfarillo}
\affiliation{CRS SMC, CNR-INFM and Dipartimento di Fisica, Universit\`a di 
Roma ``La Sapienza'', Piazzale A. Moro 2, 00185, Rome, Italy}
\author{L. Benfatto}
\affiliation{Centro Studi e Ricerche ``Enrico Fermi'', via Panisperna 89/A, 
00184, Rome, Italy} 
\affiliation{CRS SMC, CNR-INFM and Dipartimento di Fisica, Universit\`a di 
Roma ``La Sapienza'', Piazzale A. Moro 2, 00185, Rome, Italy}
\author{S. Caprara}
\affiliation{CRS SMC, CNR-INFM and Dipartimento di Fisica, Universit\`a di 
Roma ``La Sapienza'', Piazzale A. Moro 2, 00185, Rome, Italy}
\author{C. Castellani}
\affiliation{CRS SMC, CNR-INFM and Dipartimento di Fisica, Universit\`a di 
Roma ``La Sapienza'', Piazzale A. Moro 2, 00185, Rome, Italy}
\author{M. Grilli}
\affiliation{CRS SMC, CNR-INFM and Dipartimento di Fisica, Universit\`a di 
Roma ``La Sapienza'', Piazzale A. Moro 2, 00185, Rome, Italy}
\begin{abstract}
We derive the effective action for superconducting fluctuations in a
four-band model for pnictides, discussing the emergence of a single critical
mode out of a dominant interband pairing mechanism. We then apply our model to 
calculate the paraconductivity in two-dimensional and layered 
three-dimensional systems, and compare our results with recent resistivity 
measurements in SmFeAsO$_{0.8}$F$_{0.2}$.
\end{abstract}
\date{\today} 
\maketitle 

The recent discovery of superconductivity in pnictides \cite{kamihara} has
renewed the interest in high-temperature superconductivity. Pnictides share
many similarities with cuprate superconductors, e.g., the layered
structure, the proximity to a magnetic phase \cite{uemura}, the relatively
large ratio between the superconducting (SC) gap and the critical
temperature $T_c$ \cite{ding,gonnelli}, and the small superfluid density
\cite{uemura}. However, differently from cuprates, the presence of several
sheets of the Fermi surface makes the multiband character of
superconductivity an unavoidable ingredient of any theoretical model for
pnictides. Moreover, since the calculated electron-phonon coupling cannot
account for the high values of $T_c$ \cite{boeri,mazin}, it has been
suggested that the pairing glue is provided by spin fluctuations exchanged
between electrons in different bands \cite{mazin,spin_fluc,chubukov}. Thus,
pnictides are expected to be somehow different from other multiband
superconductors (e.g., MgB$_2$) where the main coupling mechanism is
intraband \cite{liu}.

This scenario raises interesting questions regarding the appropriate 
description of SC fluctuations in a multiband system dominated by interband 
pairing. The issue is relevant, because fluctuating Cooper pairs above $T_c$ 
contribute to several observable quantities, such as, e.g., the enhancement of 
dc conductivity (paraconductivity) and of the diamagnetic response as $T_c$ 
is approached \cite{LV}. The nature of SC fluctuations depends on whether the 
system is weakly or strongly coupled (and on whether preformed pairs are 
present or not) and a wealth of physical information can be obtained from 
paraconductivity and diamagnetic response, provided a theoretical background 
is established to extract them.

In this paper, after introducing a four-band model, as appropriate for 
pnictides, we discuss the subtleties related to the description of SC 
fluctuations in a system with dominant interband pairing. We then apply our 
results to compute the paraconductivity associated with SC fluctuations above 
$T_c$. We show that when interband pairing dominates, despite the presence of 
four bands, there are only two independent fluctuating modes. Only one of them 
is critical and yields a diverging Aslamazov-Larkin (AL) contribution to 
paraconductivity as $T_c$ is approached, similarly to the case of dominant 
intraband pairing \cite{varlamov}. The temperature dependence of the AL 
paraconductivity is the same derived for ordinary single-band superconductors 
\cite{al,LV}. Remarkably, within a BCS approach, we recover the AL numerical 
prefactor, which is a {\em universal} coefficient in two dimensions, and 
depends instead on the coherence length perpendicular to the planes in the 
three-dimensional layered case. We also find that subleading terms with 
respect to the leading AL contribution could distinguish between dominant 
inter- and intra-band pairing. Within this theoretical background, we analyze 
recent resistivity data in SmFeAsOF \cite{gonnelli} and discuss the results.

At present, ARPES measurements in pnictides \cite{ding} confirmed the
Fermi-surface topology predicted by LDA. It consists of two hole-like
pockets centered around the $\Gamma$ point (labeled $\a$ and $\b$,
following Ref.  [\onlinecite{ding}]), and two electron-like ($\g$) pockets
centered around the M points of the folded Brillouin zone of the FeAs
planes. Motivated by the magnetic character of the undoped parent compound
and by the approximate nesting of the hole and the electron pockets with
respect to the magnetic ordering wavevector, we assume that pairing
mediated by spin fluctuations is effective only between hole and electron
bands \cite{chubukov}. Since the $\b$ band has a Fermi surface
substantially larger than the $\a$ band, the $\b$ pocket is expected to be
less nested to the $\g$ pocket, so that the interband $\b-\g$ coupling $\l$
is smaller than the $\a-\g$ coupling $\L$, i.e.,
$\eta\equiv\lambda/\Lambda<1$. The two electron pockets have comparable
sizes and for simplicity we assume that the $\g$ bands are
degenerate. Therefore, the BCS Hamiltonian of our four-band model is
\cite{benfatto_4bands}
\be
\lb{ham}
H=\sum_i H_0^i+\Lambda \sum_\bq \left[ \Phi^\dagger_{\g,\bq}
(\Phi^{\phantom{\dagger}}_{\a,\bq}+
\eta \Phi^{\phantom{\dagger}}_{\b,\bq})+h.c.\right],
\ee
where $H_0^i=\sum_\bk
\xi_{i,\bk}c^\dagger_{i,\bk\s}c^{\phantom{\dagger}}_{i,\bk\s}$ is the band
Hamiltonian, $c_{i,\bk\s}^{(+)}$ annihilates (creates) a fermion in the
$i=\a,\b,\g_1,\g_2$ band (with the twofold degenerate $\g$ bands labeled as
$\g_1$ and $\g_2$), $\xi_{i,\bk}$ is the dispersion with respect to the
chemical potential, $\Phi_{i,\bq}=\sum_\bk
c^{\phantom{\dagger}}_{i,\bk+\bq\up} c^{\phantom{\dagger}}_{i, \bk\down}$
is the pairing operator in the $i$-th band, and
$\Phi_{\g,\bq}\equiv\Phi_{\g_1,\bq}+\Phi_{\g_2,\bq}$. Since we assumed that
pairing acts between hole and electron bands only, we can express the
pairing term in Eq.\ \ref{ham} by means of $\Phi_1\equiv\Phi_\g$ and
$\Phi_2\equiv\Phi_\a+\eta\Phi_\b$. Thus
\be
\lb{sa}
H_I=\Lambda\sum_\bq(\Phi_1^\dagger\Phi_2^{\phantom{\dagger}}+h.c.)=
-\Lambda\sum_\bq(\Phi_-^\dagger\Phi^{\phantom{\dagger}}_-
-\Phi_+^\dagger\Phi^{\phantom{\dagger}}_+),
\ee
where $\Phi_\pm=(w_1\Phi_1\pm w_2\Phi_2)/\sqrt{2}$, $w_1,w_2$ are arbitrary
numbers satisfying $w_1w_2=1$, and for definiteness we take $\Lambda>0$,
which is the case for a spin-mediated pairing interaction.  From Eq.\
\pref{sa} one immediately sees that when interband pairing dominates, the
interaction is a mixture of attraction (for $\Phi_-$) and repulsion (for
$\Phi_+$). As a consequence, when we perform the standard
Hubbard-Stratonovich (HS) decoupling of the quartic interaction term
\pref{sa} by means of the HS field $\phi$,
\bea
\lb{hs}
e^{\pm\Lambda\Phi^\dagger\Phi}&=&\int \DD \phi\, 
e^{-|\phi|^2/\Lambda+
\sqrt{\pm 1}(\Phi^\dagger\phi+h.c.)},
\eea
the HS transformation associated with the repulsive part 
contains the imaginary unit. As we shall see below, this would
require an imaginary value of the $\phi_+$ HS field at the saddle point,
that can be avoided by the rotation of the pairing operators $\Phi_\pm$
defined above, via a suitable choice of the $w_{1,2}$ coefficients.

In analogy with the single-band case \cite{lara_pp}, the effective action
for the SC fluctuations reads
\begin{eqnarray*}
S&=&\sum_i\sum_k\sum_\s(\xi_{i,\bk}-i\e_n)c^\dagger_{i,k\s}
c^{\phantom{\dagger}}_{i,k\s}\nn\\ 
&+&\sum_q\left[\frac{|\phi_+(q)|^2}{\Lambda}-i\phi_+^*(q)
\Phi^{\phantom{\dagger}}_+(q)-i\phi_+(q)\Phi^\dagger_+(q)\right]\nn\\ 
&+&\sum_q\left[\frac{|\phi_-(q)|^2}{\Lambda}-\phi_-^*(q)\Phi^{\phantom{\dagger}}_-(q)
-\phi_-(q)\Phi^\dagger_-(q)\right],~~~~\nn 
\end{eqnarray*}
where $k\equiv(\bk,i\e_n)$, $q\equiv(\bq,i\omega_m)$, and $\e_n$,
$\omega_m$ are the Matsubara fermion and boson frequencies,
respectively. Integrating out the fermions we obtain the standard
contribution to the action, $-\sum_i{\mathrm{Tr}}\log A^i_{kk'}$, where the
trace acts on momenta, frequencies, and spins. The elements of the matrices
$A^i_{kk'}$ are
\[
\left[A^i_{kk'}\right]_{11}=(\xi_{i,\bk}-i\e_n)\delta_{kk'},~~~
\left[A^i_{kk'}\right]_{22}=-(\xi_{i,\bk}+i\e_n)\delta_{kk'},
\]
with $i=\a,\b,\g_1,\g_2$,
\bea
\left[A^{\a\phantom{\b}}_{kk'}\right]_{12}&=&
\frac{w_2}{\sqrt{2}}\left[{\phi_-(k-k')-i\phi_+(k-k')}\right],\nn\\
\left[A^\b_{kk'}\right]_{12}&=&\eta\left[A^{\a\phantom{\b}}_{kk'}\right]_{12},
\nn\\
\left[A^{\g\phantom{\b}}_{kk'}\right]_{12}&=&
-\frac{w_1}{\sqrt{2}}\left[{\phi_-(k-k')+i\phi_+(k-k')}\right],\nn
\eea
for $\g=\g_1,\g_2$, and the $[A^i_{kk'}]_{21}$ elements contain the complex
conjugates of the HS fields evaluated at $(k'-k)$. 

The $q=k-k'=0$ values of the HS fields yield the SC gaps in the various
bands, within the saddle-point approximation
$\phi_\pm(q=0)=\bar\phi_\pm$. However, due to the presence of the imaginary
unit $i$ associated with the HS field $\phi_+$, in general
$[A^i_{kk}]^*_{21}\neq [A^i_{kk}]_{12}$. To recover a Hermitian $A$ matrix
at the saddle point, the integration contour of the functional integral
must be deformed toward the imaginary axis of the $\phi_+$ field.  This can
be avoided if one chooses $w_1$ and $w_2$ in the definition of $\Phi_\pm$
in such a way that $\bar\phi_+=0$. In our
case, the choice $\bar\phi_+=0$ gives $\D_\a=-w_2\bar\phi_-/\sqrt{2}$,
$\D_\g=w_1\bar\phi_-/\sqrt{2}$ and $\D_\b=\eta\Delta_\a$. Hence, the ratio
of the gaps in the two hole bands is solely determined by the ratio of the
couplings, $\D_\b/\D_\a=\eta$. Using $w_1w_2=1$ we have
$\bar\phi_-^2=-2\D_\a\D_\g$, and the equations for $\D_\a$ and $\D_\g$ read
\cite{benfatto_4bands}
\bea
\lb{eqg}
\D_\a&=&-\Lambda(2\Pi_\g)\D_\g,\\
\lb{eqa}
\D_\g&=&-\Lambda\left(\Pi_\a\D_\a+\eta\Pi_\b\D_\b\right),
\eea
where $\Pi_i=N_i\int_0^{\omega_0}d\xi[\tanh({E_i}/{2T})]/E_i$ yields the
$q=0$ value of particle-particle bubble when $T>T_c$, $N_i$ is density of
states of the $i$-th band at the Fermi level, $\omega_0$ is the cut-off for
the pairing interaction, and $E_i=\sqrt{\xi^2+\D_i^2}$. Hence, our
four-band model for pnictides effectively reduces to a two-band model, with
one electron-like and one hole-like effective band. Indeed, defining
$\D_1\equiv\D_\g$, $\D_2\equiv\D_\a$, and $\Pi_1\equiv2\Pi_\g$,
$\Pi_2\equiv\Pi_\a+\eta^2 \Pi_\b$, Eqs.\ \pref{eqg}-\pref{eqa} recover the
standard two-band expression $\D_1=-\Lambda\Pi_2\D_2$ and
$\D_2=-\Lambda\Pi_1\D_1$. Since $\D_1/\D_2=-w_1/w_2$, it also follows that
$w_1^2/w_2^2=\Pi_2/\Pi_1$ at $T<T_c$.

Let us now discuss the SC fluctuations for $T>T_c$, where $\D_{1,2}=0$. To 
derive the equivalent of the standard Gaussian Ginzburg-Landau functional 
\cite{LV}, we expand the action up to terms quadratic in the HS fields, 
$S_G=\sum_{\iota\kappa}\sum_q\phi_\iota^*(q)
L^{-1}_{\iota\kappa}(q)\phi_\kappa(q)$, where $\iota,\kappa=\pm$ and 
%
%
\be
\lb{lfluct}
L^{-1}(q)=\left(
\begin{array}{cc}
\Lambda^{-1}-\Pi_{eff}(q)&-i\,\Xi_{eff}(q) 
\\
-i\,\Xi_{eff}(q) &
\Lambda^{-1}+\Pi_{eff}(q)
\end{array}
\right),
\ee
with 
\begin{eqnarray*}
\Pi_{eff}(q)&\equiv&\frac{1}{2}[w_1^2\Pi_1(q)+w_2^2\Pi_2(q)],\nn\\ 
\Xi_{eff}(q)&\equiv&\frac{1}{2}[w_1^2\Pi_1(q)-w_2^2\Pi_2(q)].\nn
\end{eqnarray*}
The critical temperature is determined by the condition
$\mathrm{det}\,L^{-1}(q=0)=0$. In the BCS approximation
$\Pi_i(0)=N_i\ln(1.13\,\omega_0/T_c)$ and, in agreement with the Eqs.\
\pref{eqg}-\pref{eqa}, we obtain
$N_{eff}\Lambda\log(1.13\,\omega_0/T_c)=1$. Here, the parameter
$N_{eff}\equiv\sqrt{N_1 N_2}=\sqrt{2N_\g(N_\a+\eta^2 N_\b)}$ plays the role
of an effective density of states. To compute the fluctuation contribution
to the various physical quantities at $T>T_c$, we evaluate Eq.\
\pref{lfluct} at leading order in $q$ (hydrodynamic approximation), using
the standard expansion of the particle-particle bubble for a layered
system,
$\Pi_i(q)\approx\Pi_i(0)-c_{i,\parallel}q^2_\parallel-c_{i,z}q_z^2-\g_i|\omega_m|$,
with $q^2_\parallel=q_x^2+q_y^2$ and the $z$ axis perpendicular to the
layers.  In the BCS approximation, e.g., $\g_i=\pi N_i/(8T)$. We omit the
lengthier but standard BCS expressions for $c_{i,\parallel}$ and $c_{i,z}$
(see, e.g., Ref. [\onlinecite{LV}]), that will not be explicitly used in
the following.  By making for $T>T_c$ the same choice previously adopted
for $T<T_c$, $w_1^2\Pi_1(0)=w_2^2\Pi_2(0)$, we simplify the structure of
the fluctuating modes. Indeed, with this choice, the off-diagonal terms in
Eq.\ \pref{lfluct} yield contributions beyond hydrodynamics and can be
neglected. Therefore, the leading SC fluctuations are described by
\be
\lb{hydro}
L^{-1}(q)\approx\left(
\begin{array}{cc}
m_-+\n(q)&0 \\
0 & m_+-\n(q)
\end{array}
\right),
\ee
where $m_\pm=\Lambda^{-1}\pm\sqrt{\Pi_1\Pi_2}$ are the masses of the
collective modes, $\n(q)\equiv c_\parallel q_\parallel^2+c_zq_z^2+\g
|\omega_m|$, $c_\parallel=(w_1^2c_{1,\parallel}+w_2^2c_{2,\parallel})/2$ is
the stiffness along the layers, $c_z=(w_1^2c_{1,z}+w_2^2c_{2,z})/2$ is the
stiffness in the direction perpendicular to the layers, and
$\g=(w_1^2\g_1+w_2^2\g_2)/2$ is the damping coefficient.  In the BCS case
$m_\pm=N_{eff}[\ln(1.13\,\omega_0/T_c)\pm \ln(1.13\,\omega_0/T)]$ and
$\gamma=\pi N_{eff}/(8T)$.

Having deduced the hydrodynamic action of the SC fluctuations, we can 
calculate the paraconductivity along the lines of Ref.\ 
[\onlinecite{caprara}]. By inspection of Eq.\ \pref{hydro}, one can see that 
only the $\phi_-$ mode becomes critical at $T_c$ [i.e., $m_-(T_c)=0$,
$m_+(T_c)=2\Lambda^{-1}$], thus giving a diverging fluctuation contribution to 
various physical quantities when $T\ra T_c$. The leading contribution to the 
current-current response function along the layers is 
\[
\delta\chi(\Omega_\ell)=4e^2T\sum_q c_\parallel^2 q_\parallel^2 L_{--}
(\bq,\omega_m)
L_{--}(\bq,\omega_m+\Omega_\ell),
\]
whence the paraconductivity
$\delta\sigma_{AL}=[{\mathrm{Im}}\delta\chi(\Omega)/\Omega]_{\Omega\ra 0}$
is obtained, after analytical continuation of the Matsubara frequency
$\Omega_\ell$ to the real frequency $\Omega$. Therefore, the same
expressions known for a single-band superconductor are found, although with
the effective parameters $m_-,c_z,\g$.  The paraconductivity along the
layers is independent of the in-plane stiffness $c_\parallel$, as
guaranteed by the same gauge-invariance arguments discussed in Ref.\
[\onlinecite{caprara}] for a single-band superconductor. The leading
contributions to paraconductivity along the layers in three and two
dimensions (3D and 2D, respectively) take the AL form \cite{al}
\bea
\lb{3d}
\d\s_{AL}^{3D}&=&\frac{e^2}{32\hbar\xi_z}\frac{1}{\sqrt{\epsilon}}\\
\lb{2d}
\d\s_{AL}^{2D}&=&\frac{e^2}{16\hbar d}\frac{1}{\epsilon}
\eea
where $\xi_z=\sqrt{c_z/\g}$ is the correlation length in the direction
perpendicular to the layers, $d$ is the distance between layers, and
$\epsilon(T)\equiv\pi m_-(T)/(8\g T_c)$ is the dimensionless mass of the
critical collective mode. We point out that the above expressions are
general within a hydrodynamic description of the collective modes and do
not rely on any particular assumption about the pairing strength
\cite{caprara}. When the BCS expression for the $\Pi_i$ bubbles holds,
$m_-(T)=N_{eff}\ln(T/T_c)$ and the dimensionless mass appearing in
Eqs. \pref{3d}-\pref{2d} is simply $\epsilon=\log(T/T_c)$.

The calculated AL paraconductivity expressions \pref{3d}-\pref{2d} can be now
compared with the existing results for the pnictides. As it has been
widely discussed in the context of cuprates \cite{caprara05}, for
weakly-coupled layered materials the SC fluctuations usually display a
2D-3D dimensional crossover as $T_c$ is approached. However, the interlayer
coupling has a different relevance in the various families of cuprates,
with substantial 3D behavior [Eq. (\ref{3d})] in \ybco\ samples, while more
anisotropic \bscco\ or \lasco\ compounds show 2D fluctuations [Eq.
(\ref{2d})], the 2D-3D crossover being too close to $T_c$ to be clearly
observed. Such a systematic survey has not yet been performed in the case
of pnictides, due also to the limited availability of clean single crystal.
Indeed, the analysis of the 2D-3D crossover might be biased in polycrystals
by the distribution of critical temperatures and by the mixing of the
planar and perpendicular directions.

\begin{figure}
\includegraphics[scale=0.3]{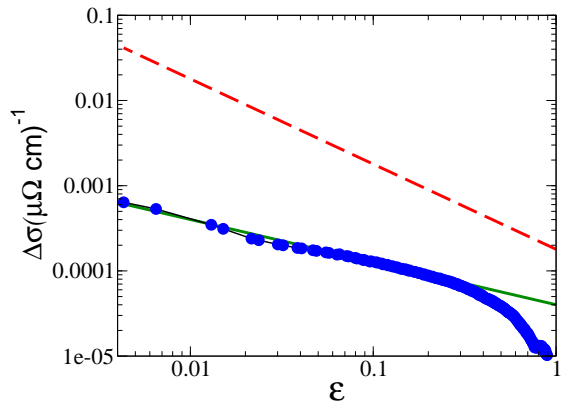}
\caption{(Color online) Comparison between the experimental paraconductivity 
for a SmFeAsO$_{0.8}$F$_{0.2}$ sample studied in Ref. [\onlinecite{gonnelli}] 
(black circles) and the 2D (dashed line) and 3D (solid line) expressions of 
the AL paraconductivity [Eqs. (\ref{3d},\ref{2d})]. For the 3D 
paraconductivity a coherence length $\xi_z=19$\,\AA\ has been used, while for 
the 2D case the structural distance between layers $d=8.4$\,\AA\ has been 
inserted.}
\label{fig1}
%
\end{figure}

Having in mind such limitations, we attempt the analysis of paraconductivity 
in a SmFeAsO$_{0.8}$F$_{0.2}$ sample \cite{gonnelli} with $T_c\approx 52$\,K. 
To determine the contribution of SC fluctuations to the normal-state 
conductivity, $\delta \sigma\equiv \rho^{-1} -\rho_N^{-1}$, we need to extract
the normal state resistivity $\rho_N$ from the data. Owing to the 
diverging conductivity, the precise determination of the normal-state 
contribution is immaterial near $T_c$, but becomes relevant for larger values 
of $\epsilon$. We fitted the resistivity at high temperature (in the range 
between $279$\,K, the highest available temperature, and about $200$\,K) 
checking that the qualitative results were stable upon small variations of 
this range. We used a quadratic fit $\rho_N=a+b(T-T_0)+c(T-T_0)^2$, with 
$T_0=279$\,K, and found that the resulting paraconductivity is roughly two 
orders of magnitude smaller than the 2D AL result [Eq. (\ref{2d})]. The 
slope in a log-log plot is in agreement with the 3D power law 
[Eq. (\ref{3d})]. The plot in Fig.\ \ref{fig1} clearly shows that the 3D 
behavior extends over two decades of $\epsilon$. The fitting parameters 
$a=1700$\,$\mu\Omega$\,cm, $b=21.9$\,$\mu\Omega$\,cm/K, and 
$c=0.018$\,$\mu\Omega$\,cm/K$^2$ were used. 

The fitting with Eq. (\ref{3d}) allows us to determine the precise value of
$T_c$ and the coherence length $\xi_z$ of the SC fluctuations in the
direction perpendicular to the FeAs layers. We find $T_c=51.4$\,K and
$\xi_z=19$\,\AA.  One can also see that substantial SC fluctuations persist
up to $\epsilon\approx (T-T_c)/T_c\approx 0.4$, i.e., up to $20$\,K above
$T_c$.  This fluctuating regime is therefore much larger than in
conventional 3D superconductors and, despite the 3D behavior of the
paraconductivity, calls for a relevant character of the layered structure
and for a small planar coherence length. We point out that at even larger
values of $\epsilon$ the paraconductivity drastically drops, in analogy
with what found in cuprates (see, e.g., Ref. [\onlinecite{caprara05}], and
references therein). The way paraconductivity deviates from the AL behavior
in multiband systems also depends on the role of the other (non-critical)
collective modes. In particular, it can be shown \cite{future} that, when
the intraband pairing is equally dominant in all bands, the
paraconductivity mediated by the non critical modes may become sizable, and
the experimental data should approach the pure AL contribution of the
critical mode from above. This is not the case when the dominant pairing is
interband and therefore it is not surprising that the data for the pnictide
sample analyzed in this paper always lay below the AL straight line in
Fig.\ 1.

In conclusion, we investigated the occurrence of SC fluctuations in a
multiband system where interband pairing dominates, as appropriate for
pnictides. In contrast to the case of dominant intraband mechanism (as,
e.g., in MgB$_2$ \cite{varlamov}), in the present situation the HS
decoupling must be accompanied with a proper rotation of the fermion fields
which guarantees a Hermitian saddle-point action below $T_c$. The same
rotation leads to a straightforward decoupling of the Gaussian fluctuations
above $T_c$ in the hydrodynamic limit. Thus, despite the apparent
complexity of the multiband structure in pnictides, we demonstrate that the
AL expressions for paraconductivity stay valid not only as far as the
functional temperature dependence is concerned, but also regarding the
numerical prefactors. While in the BCS 2D case the prefactor stays
universal, in the 3D case the only difference is that a suitable
redefinition of the transverse coherence length has to be introduced.  With
this equipment, we considered the experimental resistivity data of the
SmFeAsO$_{0.8}$F$_{0.2}$ sample studied in Ref. [\onlinecite{gonnelli}],
finding that here fluctuations have a 3D character and extend far above
$T_c$. Recently, an experimental work\cite{marina} on fluctuation
conductivity in pnictides confirmed the wide fluctuating regime, even
though fluctuations seem to have 2D character. Thus, further experiments
are in order to confirm the nature of fluctuations in pnictides, and to
assess the relevance of Cooper-pair fluctuations in these new
superconductors.

{\it Acknowledgments.} We warmly thank D.~Daghero and R.~Gonnelli for
providing us with the data of Ref.\ \cite{gonnelli}. This work has been
supported by PRIN 2007 (Project No. 2007FW3MJX003).


\begin{thebibliography}{99}
\bibitem{kamihara} 
Y.~Kamihara, T.~Watanabe, M.~Hirano, and H.~Hosono, J. Am. Chem. Soc.
{\bf 130}, 3296 (2008).

\bibitem{uemura} See, e.g.,
T.~Goko, {\em et al.}
arXiv:0808.1425.

\bibitem{ding} H.~Ding, P.~Richard, K.~Nakayama, K.~Sugawara, T.~Arakane, 
Y.~Sekiba, A.~Takayama, S.~Souma, T.~Sato, T.~Takahashi, Z.~Wang, X.~Dai, 
Z.~Fang, G.~F.~Chen, J.~L.~Luo and N.~L.~Wang, Europhys. Lett. {\bf 83}, 47001 
(2008).

\bibitem{gonnelli}
D.~Daghero, M.~Tortello, R.~S.~Gonnelli, V.~A.~Stepanov, N.~D.~Zhigadlo, 
J.~Karpinski, arXiv:0812.1141.

\bibitem{boeri} L.~Boeri, O.~V.~Dolgov, and A.~A.~Golubev, Phys. Rev. Lett. 
{\bf 101}, 026403 (2008)

\bibitem{mazin} I.~I.~Mazin, D.~J.~Singh, M.~D.~Johannes, M.~H.~Du,
Phys. Rev.  Lett. {\bf 101}, 057003 (2008); 

\bibitem{spin_fluc} 
V.~Stanev, J.~Kang and Z.~Tesanovic, \prb {\bf 78}, 184509 
(2008); R.~Sknepnek, G.~Samolyuk, Y.~Lee, and J.~Schmalian, 
\prb {\bf 79}, 054511 (2009).

\bibitem{chubukov}
 A.~V.~Chubukov, D.~Efremov, I.~Eremin, \prb {\bf 78}, 134512
(2008); V.~Cvetkovic and Z.~Tesanovic, Europhys. Lett. {\bf 85}, 37002
 (2009).  

\bibitem{liu}
A.~Y.~Liu, I.~Mazin, and J.~Kortus, Phys. Rev. Lett. {\bf 87}, 087005
(2001).

\bibitem{LV} A. Larkin and A. Varlamov, \textit{Theory of fluctuations in 
superconductors}, (Clarendon Press, Oxford, 2005).

\bibitem{varlamov}
A.~E.~Koshelev, A.~A.~Varlamov and V.~M.~Vinokur, \prb {\bf 72}, 064523
(2005). 


\bibitem{al} L. G. Aslamazov and A. I. Larkin, Phys. Lett. A \textbf{26}, 238 
(1968); Sov. Phys. Solid State \textbf{10}, 875 (1968).

\bibitem{benfatto_4bands}
L.~Benfatto, M.~Capone, S.~Caprara, C.~Castellani, and C. Di Castro, \prb 
{\bf 78}, 140502(R) (2008).

\bibitem{lara_pp}
For a list of references in the single-band case see, e.g., 
L.~Benfatto, A.~Toschi, and S.~Caprara, Phys. Rev. B. {\bf 69}, 
184510 (2004).

\bibitem{caprara} S. Caprara, M. Grilli, B. Leridon, and J. Vanacken, Phys. 
Rev. B {\bf 79}, 024506 (2009).

\bibitem{caprara05} S. Caprara, M. Grilli, B. Leridon, and J. Lesueur, Phys. 
Rev. B {\bf 72}, 107509 (2005).

\bibitem{future} L. Benfatto, S. Caprara, C. Castellani, L. Fanfarillo, and M. 
Grilli, unpublished.

\bibitem{marina}
I. Pallecchi, C. Fanciulli, M. Tropeano, A. Palenzona, M. Ferretti,
A. Malagoli, A. Martinelli, I. Sheikin, M. Putti, and C. Ferdeghini, \prb
{\bf 79}, 104515 (2009).

\end{thebibliography}
\end{document}